\documentclass[amsmath,amssymb,floatfix,aps,showpacs,superscriptaddress,nofootinbib,notitlepage]{revtex4-1}

\usepackage{amsmath,amsthm,amssymb,bm,hyperref,graphicx,color,float,changes,mathtools}
\allowdisplaybreaks

\newtheorem{lem}{Lemma}[section]

\newcommand{\be}{\begin{equation}}
\newcommand{\ee}{\end{equation}}
\newcommand{\bea}{\begin{eqnarray}}
\newcommand{\eea}{\end{eqnarray}}
\newcommand{\6}{\partial}

\newcommand{\inti}{\int_{-\infty}^{+\infty}}

\begin{document}

\title{Universal Tan relations for quantum gases in one dimension}

\author{Ovidiu I. P\^{a}\c{t}u}
\affiliation{Institute for Space Sciences, Bucharest-M\u{a}gurele, R 077125, Romania}
\author{Andreas Kl\"umper}
\affiliation{Fakult\"at f\"ur Mathematik und
    Naturwissenschaften, Bergische Universit\"at Wuppertal, 42097 Wuppertal,
    Germany}

\begin{abstract}

We investigate universal properties of one-dimensional multi-component systems
comprised of fermions, bosons, or an arbitrary mixture, with contact
interactions and subjected to an external potential. The masses and the
coupling strengths between different types of particles are allowed to be
different and we also take into account the presence of an arbitrary magnetic
field. We show that the momentum distribution of these systems exhibits
a universal $n_\sigma(k) \sim C_\sigma/k^4$ decay with
$C_\sigma$ the contact of species $\sigma$ which can be computed from the
derivatives of an appropriate thermodynamic potential with respect to the
scattering lengths. In the case of integrable fermionic systems we argue that
at fixed density and repulsive interactions the total contact reaches its
maximum in the balanced system and monotonically decreases to zero as we
increase the magnetic field. The converse effect is present in integrable
bosonic systems: the contact is largest in the fully polarized state and
reaches its minimum when all states are equally populated. We obtain short
distance expansions for the Green's function and pair distribution function
and show that the coefficients of these expansions can be expressed in terms
of the density, kinetic energy and contact. In addition we derive universal
thermodynamic identities relating the total energy of the system, pressure,
trapping energy and contact. Our results are valid at zero and finite
temperature, for homogeneous or trapped systems and for few-body or many-body
states.
\end{abstract}


\maketitle

\onecolumngrid

\section{Introduction}

Many current experiments with ultracold atomic gases investigate the so-called
zero range regime in which the thermal de Broglie wavelength and the average
interparticle distance is much larger than the effective range of the
interaction potential. In this regime the thermodynamic properties of the gas
are universal and depend on the interaction potential through the s-wave
scattering length which characterizes the low-energy scattering of two
atoms. In the case of 3D spin-$\frac{1}{2}$ fermions in the zero range regime
Tan has shown \cite{Tan1,Tan2,Tan3} that the large momentum
distribution exhibits a universal decay $n_{\uparrow,\downarrow}(k)\sim C/k^4$
with $C$ an extensive quantity called contact which is the same for spin-up
and -down particles. In 1D a similar decay of the momentum distribution for
the Lieb-Liniger model \cite{LL} (bosons with delta-function interaction) was
derived earlier for any value of the interaction strength by Olshanii and
Dunjko in \cite{OD}. The contact, which in the case of spin-$\frac{1}{2}$
fermions quantifies the probability that two particles of opposite spin to be
found in the same region of space, appears in a series of relations, known
also as Tan relations, connecting the short distance expansions of the one-
and two-particle reduced density matrices and thermodynamic properties such as
the energy and pressure. While initial theoretical investigations were focused
on 3D spin-$\frac{1}{2}$ fermions \cite{Tan1,Tan2,Tan3, BP1,BKP,ZL,CAL} they
were soon extended in the case of 2D systems \cite{WC1,VZM1,
VZM2}, 3D bosonic systems \cite{WC2}, general 2D and 3D
mixtures \cite{WC2} and 1D spin-$\frac{1}{2}$ fermions \cite{BZ}. These
powerful universal relations are important and interesting not only because
they relate a microscopic quantity to macroscopic quantities like the energy,
but also due to their wide range of applicability being valid both in the
superfluid and normal phase, at zero or finite temperature in the case of
few-body or many-body states. The contact is also experimentally accessible
through a variety of methods that measure: a) the molecular fraction from
photo-association experiments \cite{WTC,PSKJH}, b) the tail of the momentum
distribution \cite{SGDJ}, c) the RF spectroscopy signal at large frequencies
\cite{SGDJ,WMPCJ,SDPJ}, d) the equation of state \cite{NNCS} e) and the static
structure factor at large momenta using Bragg spectroscopy \cite{KHLDV1,KHHDH,
 KHDHH,HLFHV}. In addition to the measurement of the contact several Tan
relations were also checked experimentally \cite{SGDJ,KHLDV1,KHHDH,KHDHH}.

In 1D the $1/k^4$ decay of the momentum distribution and the contact
were investigated numerically and analytically for the  Lieb-Liniger model
\cite{L1,VT1,VT2,JMMS,MVT,LGW,FFGW,CCS,V,VM,WHLYGB,XR,LVM,STN}, attractive and
repulsive spin-$\frac{1}{2}$ fermions
\cite{CSZ,GM,CJGZ,VFJVZ,LKTZ,BAD,HJLPAD,RPLD,LBDR,HCZG,MSPD,PK}, the fermionic polaron
problem \cite{DK1,DK2,DKTK,LKTZ}, spin-$\frac{1}{2}$ bosons \cite{CSZ,BBF,PKF},
spin-1 bosons \cite{JY1,JY2}, $\kappa$-component ($\kappa>2$) fermions and
mixtures \cite{GMW,HLPD,RPBD,DJARMV1, DJARMV2}, and impenetrable anyons
\cite{SSC,SC,HS}. Despite this intense activity a comprehensive set of Tan
relations or an analytical derivation of the tail of the momentum distribution
can be found in the literature only for the case of single-component bosons
\cite{OD,V,STN} and balanced spin-$\frac{1}{2}$ fermions \cite{BZ}. Further
motivated by the recent experimental realization \cite{Pag} of 1D
multi-component repulsive fermions (with $\kappa=2,\cdots,6$), in this article
we derive Tan relations for the most general multi-component 1D system with
contact interactions in an external potential. In \cite{Tan1,Tan2,Tan3} Tan
used a new type of generalized functions to prove the universal relations
involving the contact.  Subsequently, it was shown that they can be rederived
using the Operator Product Expansion of local operators \cite{BP1,BKP}. Here,
we use what we consider to be the most powerful and transparent technique
which uses the fact that the derivative of the wave-function of a system with
contact interactions is discontinuous when the coordinates of two particles
coincide \cite{BR1,BR2}. This is the same method employed by Olshanii and
Dunjko \cite{OD} in their study of the short distance, large momentum
distribution of the Lieb-Liniger model and by Werner and Castin in their
comprehensive papers on Tan relations for 2D and 3D mixtures \cite{WC1,WC2}.
First, we establish that the derivative of the wave-function has a
discontinuity which is a direct consequence of the delta function
interaction. This discontinuity produces a $1/k^2$ large wave-number
asymptotics of the wave-function in the momentum representation which implies
that the momentum distribution (which is bilinear in the momentum
representation of the wave-function) presents a $1/k^4$
tail. Then, the short-distance expansion of the Green's function can be
obtained from the inverse Fourier transform of the momentum distribution and
identification of certain terms with known quantities.

The plan of the paper is as follows. In Sect.~\ref{S2} we introduce the model
and the definition of the relevant quantities. In Sect.~\ref{S3} we study the
large momentum distribution and we determine the contact. Short distance
expansion of the correlation functions and thermodynamic identities are
derived in Sects. \ref{S4} and \ref{S5}. The particular case of integrable
models is analyzed in Sect.~\ref{S6}.  We conclude in Sect.~\ref{S7}.

\section{Model and correlation functions}\label{S2}

We consider a one-dimensional system comprised of $\kappa$-types of particles interacting via a
delta-function potential in an external potential. The particles can be  bosons, fermions or an arbitrary
mixture. Different types of particles will be labeled by an integer $\sigma\in\{1,\cdots,\kappa\}$ and the
number of particles of each species will be denoted by $N_\sigma$ with the total number of particles being
$N=\sum_\sigma N_\sigma$. In second quantization the Hamiltonian is $H=\int \mathcal{H}(x)\, dx$ with the
Hamiltonian density
\be\label{ham}
\mathcal{H}(x)= \sum_\sigma\frac{\hbar^2}{2m_\sigma}\6_x\Psi^\dagger_\sigma(x)\6_x\Psi_\sigma(x)
+\sum_{\sigma \le \sigma'}\frac{g_{\sigma\sigma'}}{1+\delta_{\sigma\sigma'}}
\Psi^\dagger_\sigma(x)\Psi^\dagger_{\sigma'}(x)\Psi_{\sigma'}(x)\Psi_\sigma(x)
+\sum_\sigma\left(V_\sigma(x)-\mu_\sigma\right)\Psi^\dagger_\sigma(x)\Psi_\sigma(x)\, .
\ee
Here, $\Psi^\dagger_\sigma(x)$ and $ \Psi_\sigma(x)$ are bosonic and/or fermionic fields which obey
canonical commutation or anticommutation relations, $m_\sigma$ and $\mu_\sigma$ are the mass and the
chemical potentials of particles of type $\sigma$  and $V_\sigma(x)=m_\sigma \alpha_\nu x^\nu$
is the external trapping potential with $\nu=2,4,\cdots.$
The presence of an external magnetic field is taken into account by allowing for the chemical potentials of
different species to be different. The coupling strengths can be expressed in terms
of the  1D scattering length $a_{\sigma\sigma'}$ via $g_{\sigma\sigma'}=-\hbar^2/ (m_{\sigma\sigma'}
a_{\sigma\sigma'})$  where $m_{\sigma\sigma'}=m_\sigma m_{\sigma'}/(m_\sigma+m_{\sigma'})$
is the reduced mass. We will consider attractive and repulsive values of the coupling strength in the
case of fermionic particles and only repulsive interactions between bosonic particles. Due to the fact
that  $a_{\sigma\sigma'}=a_{\sigma'\sigma}$  there are only  $\kappa(\kappa+1)/2$ distinct
scattering lengths.

In order to study correlation functions we need to introduce some notations. We will say that particle
$i$ is of type $\sigma$  if $i\in I_\sigma$ where $I_\sigma$ is a partition of $\{1,\cdots, N\}$ in
$\kappa$ subsets with cardinalities $N_\sigma$. The most obvious choice, which will be used throughout
this paper, is $I_1=\{1,\cdots,N_1\}\, ,$ $I_2=\{N_1+1,\cdots, N_1+N_2\}\, , \cdots$ which means that
the first $N_1$ particles are of type  $1$ the next $N_2$ particles are of type $2$, etc. We will denote
by $P_{\sigma\sigma'}$ the set of all pairs of particles with  one particle in group $\sigma$ and the
other one in $\sigma'$ counted only once i.e., $P_{\sigma\sigma'}\equiv \{(i,j)\in (I_\sigma\times
I_{\sigma'})\, /\, i<j \}$. For example, if $\kappa=2$  $N=5$, $I_1=\{1,2,3\}$  and $I_2=\{4,5\}$ then $P_{1,2}=
\{(1,4),\, (1,5),\, (2,4),\, (2,5),\, (3,4),\, (3,5)\}$. Using these notations the Hamiltonian
(\ref{ham}) in first quantized form can be written as
\be\label{ham2}
H=-\sum_\sigma\sum_{i\in I_\sigma}\frac{\hbar^2}{2 m_\sigma}\frac{\6^2}{\6 x_i^2}
+ \sum_{\sigma\le\sigma'}\sum_{(i,j)\in P_{\sigma\sigma'}}g_{\sigma\sigma'}\delta(x_i-x_j)
+\sum_\sigma\sum_{i\in I_\sigma}V_\sigma(x_i)-\sum_\sigma\mu_\sigma N_\sigma\, .
\ee

A consequence of the delta function appearing in Hamiltonian (\ref{ham2}) is
that the derivative of the wave-function is discontinuous when the coordinates
of any two particles coincide and regular elsewhere. More precisely, if we
consider the limit when the relative distance between two particles of species
$\sigma$ and $\sigma'$ becomes zero and all the other coordinates are kept
fixed we have (see Appendix \ref{a1})
\be\label{disc}
\lim_{x_i\rightarrow x_j}\psi_N(x_1,\cdots,x_i,\cdots,x_j,\cdots,x_N)\sim \psi_N(x_1,\cdots, X_{ij},\cdots,
X_{ij},\cdots,x_N) \left[1-|x_{ij}|/a_{\sigma\sigma'}+\mathcal{O}(x_{ij}^2)\right]\, ,
\ee
where $X_{ij}= (m_\sigma x_i+m_{\sigma'}x_j)/(m_\sigma+m_{\sigma'})$ and  $x_{ij}=x_i-x_j$ are the
center of mass and relative coordinates of the two particles. Due to the discontinuity the Fourier
transform of the wave-functions at large momenta will have an $1/k^2$  asymptotic behavior \cite{BR1,BR2}.
This fundamental observation  will be used extensively throughout this paper and will constitute our main
tool in deriving the Tan relations.

The Green's function or the one-body density matrix is defined as

\be\label{green}
g_{\sigma\sigma}^{(1)}(x,x')\equiv\langle \Psi_\sigma^\dagger(x)\Psi_\sigma(x')\rangle\, \\
=\sum_{i\in I_\sigma}\int\prod_{\substack{j=1,\, j\ne i}}^N dx_j\,
\psi^*_N(x_1,\cdots,\underset{\underset{i}{\uparrow}}{x},\cdots,x_N)\psi_N(x_1,\cdots,\underset{\underset{i}
{\uparrow}}{x'},\cdots,x_N)\, ,
\ee
with the arrows showing that $x$ and $x'$ appear on the $i$-th position.  If
we assume that the wave-functions are normalized to one $\int \prod_{j=1}^N
dx_j\, |\psi_N(x_1,\cdots,x_N)|^2=1$ then we have $\int
g_{\sigma\sigma}^{(1)}(x,x)\, dx=N_\sigma$. The momentum distribution is the
Fourier transform of the one-body density matrix
\begin{align}
n_\sigma(k)&=\iint\, e^{-ik(x-x')}g_{\sigma\sigma}^{(1)}(x,x')\,dx\,dx'\,
=\sum_{i\in I_\sigma}\int\prod_{\substack{j=1,\, j\ne i}}^N dx_j\left|\int e^{-i k x}
\psi_N(x_1,\cdots,\underset{\underset{i}{\uparrow}}{x}
,\cdots,x_N)\, dx\right|^2\, ,\label{mom}
\end{align}
where in the last relation we have used the definition (\ref{green}) and interchanged the order of integration.
Using the integral representation of the delta function, $\int e^{-ikx}\, dk=2\pi \delta(x-x')$ we obtain
the normalization of the momentum distribution  $ \int n_\sigma(k)\, dk=2\pi\int
g_{\sigma\sigma}^{(1)}(x,x)\,dx=2\pi N_\sigma\, .
$

The pair distribution function is defined as $g_{\sigma\sigma'}^{(2)}(x,x')\equiv\langle \Psi_\sigma^\dagger(x)
\Psi_{\sigma'}^\dagger(x')\Psi_{\sigma'}(x') \Psi_\sigma(x)\rangle$ and has the following expression in
terms of the wave-function
%
\begin{align}\label{pair}
g_{\sigma\sigma'}^{(2)}(x,x')
=\int\prod_{\substack{l=1}}^N dx_l\,
\left| \psi_N(x_1,\cdots,x_N)\right|^2\sum_{i\in I_\sigma,\,  j\in I_{\sigma'},\, i\ne j }\delta(x-x_i)
\delta(x'-x_j)\, .
\end{align}
It is important to note that  $g_{\sigma\sigma'}^{(2)}(x,x')=g_{\sigma'\sigma}^{(2)}(x',x)$. Under the
assumption of normalization of the wave-functions  we have $\int g_{\sigma\sigma}^{(2)}(x,x')\, dxdx'
=N_{\sigma}(N_{\sigma}-1)$ and $\int g_{\sigma\sigma'}^{(2)}(x,x')\, dxdx'=N_{\sigma} N_{\sigma'}$.

\section{Tail of the momentum distribution and expression for the contact}\label{S3}

We start our investigation by showing that for all systems described by the
Hamiltonians (\ref{ham}) the large momentum distribution has an
$n_\sigma(k)\sim C_\sigma/k^4$ decay with $C_\sigma$ an extensive quantity
called contact. The contact can be expressed in terms of the pair distribution
functions $g_{\sigma\sigma'}^{(2)}(x,x)$ and can be computed
from the thermodynamic properties of the system via the Helmann-Feynman
theorem.  Our proof will follow the original idea of Olshanii and Dunjko
\cite{OD} first employed in the case of the Lieb-Liniger model and
subsequently generalized by Werner and Castin \cite{WC1,WC2} for 2D and 3D
systems. For reasons of clarity we will first remind of the calculations for
single component bosons and then show how the same method can be used in the
case of two-component systems.  As we will see the general result can be
easily inferred form these particular cases.

\subsection{Warm up. The Lieb-Liniger model}

The simplest realization of the Hamiltonian (\ref{ham}) is in terms of single component bosons which
is also known as the Lieb-Liniger model (single component fermions are equivalent with free fermions
because spin polarized fermions do not ``feel" the contact interaction). The homogeneous Lieb-Liniger
model is  integrable and for the last 60 years its correlation functions have been the subject of extensive
theoretical and numerical investigations. The Hamiltonian density has the following simple form
\be\label{hamLL}
\mathcal{H}(x)= \frac{\hbar^2}{2m}\6_x\Psi^\dagger(x)\6_x\Psi(x)
+\frac{g}{2} \Psi^\dagger(x)\Psi^\dagger(x)\Psi(x)\Psi(x)+(V(x)-\mu)\Psi^\dagger(x)\Psi(x)\, ,
\ee
where $\Psi^\dagger(x), \Psi(x)$ are canonical bosonic fields,  $m$ is the mass of the particles and
$g=- 2\hbar^2/m a.$   The short distance expansion of the one-body density matrix
of impenetrable bosons was first obtained by Lenard \cite{L1} (see also \cite{VT1,VT2,JMMS}) who showed
that the first nonanalytic contribution appears in the $|x|^3$ term (we will show below that this implies
a $1/k^4$ decay  of the momentum distribution). In the case of trapped impenetrable bosons the $1/k^4$
decay was investigated in \cite{MVT,VM} and for any finite interaction it was derived in \cite{OD}
(numerical results for various finite couplings can be found in \cite{CCS}). Below, we follow \cite{OD}.

We are interested in computing the decay of the momentum distribution at large momenta. Eq.~(\ref{mom})
shows that this task is accomplished by investigating the large-$k$ asymptotic behavior of the Fourier
transform of the  wave-function. In this limit the main contributions will be given by the points where
the derivative of the wave-function is discontinuous (\ref{disc}) or, more precisely, when the coordinates
of two particles are equal. We should stress that $\psi_N(x_1,\cdots, x_i,\cdots,x_i,\cdots,x_N)$
is regular if all the other variables except $x_i$ are different. Our analysis will use the following
result:
\begin{lem}\label{lemma}\cite{BH}
Consider a function absolutely integrable, vanishing at infinity which has a
singularity of the type $f(x)=|x-x_0|^\alpha F(x)$ at $x_0$ with $F(x)$
analytic and $\alpha>-1\, , \alpha\ne 0,2,4,\cdots$. Then, we have
\be\label{asympt} \lim_{k\rightarrow\infty}\inti e^{- i k x} f(x)\, dx\sim
2\cos \frac{\pi}{2}(\alpha+1)\Gamma(\alpha+1) \frac{ e^{-i k
    x_0}}{|k|^{\alpha+1}}F(x_0)+\mathcal{O}(1/|k|^{\alpha+2})\, .  \ee In the
case of multiple singular points of the same type the asymptotic behaviour of
the integral is given by the sum of all the corresponding contributions given
by the r.h.s of (\ref{asympt}).
\end{lem}

Now, we have all the tools to complete our analysis. First, we notice that as a result of the bosonic
symmetry Eq.~(\ref{mom}) can be rewritten as (in this case $I_\sigma=\{1,\cdots,N\}$),
$
n(k)=N \int\prod_{\substack{j=2}}^N dx_j\left|\int e^{-i k x_1}\psi_N(x_1,x_2,\cdots,x_N)\, dx_1\right|^2\, ,
$
which shows that we can focus our analysis on the Fourier transform with respect to the first particle.
Taking into account (\ref{disc}) and employing Lemma \ref{lemma} at the points $x_1=\{x_2,\cdots,x_N\}$
we find

\begin{align*}
\lim_{k\rightarrow\infty}\int e^{-i k x_1}\psi_N(x_1,x_2,\cdots,x_N)\, dx_1\, &\sim
\int e^{-i k x_1}\sum_{j=2}^N \psi_N(\underset{\underset{1}{\uparrow}}{X_{1j}},\cdots,
\underset{\underset{j}{\uparrow}}{X_{1j}},\cdots,x_N)\left[1-\frac{|x_{1j}|}{a}+\mathcal{O}
(|x_{1j}|^2)\right]\, dx_1\, ,\\
&\sim
\sum_{j=2}^N \frac{ e^{-i k x_j}}{k^2}\frac{2}{a}\psi_N(x_j,\cdots,x_j,\cdots,x_N)\, .
\end{align*}
where we have used  $\lim_{x_1\rightarrow x_j}X_{1j}=x_j$. Using this last relation in the expression
for the momentum distribution we obtain
\begin{align}
\lim_{k\rightarrow\infty}n(k)&\sim N\int \prod_{m=2}^N dx_m\, \sum_{j=2}^N\sum_{l=2}^N
\frac{4 e^{-i k(x_j-x_l)}}{k^4 a^2 }
\psi_N(x_j,\cdots,x_j,\cdots,x_N)\psi_N^*(x_l,\cdots,x_l,\cdots,x_N)\, ,\nonumber\\
&\sim \frac{4}{k^4}\frac{N}{ a^2 } \sum_{j=2}^N\int \prod_{i=2}^N dx_i\,
|\psi_N(x_j,\cdots,x_j,\cdots,x_N)|^2\,
= \frac{4}{k^4}\frac{1}{ a^2 }\int g^{(2)}(x,x)\, dx \, . \label{i2}
\end{align}
Here, the second line is obtained by neglecting the off-diagonal terms which vanish faster
than $1/k^4$ due to the Riemann-Lebesgue lemma  and using the definition (\ref{pair}) of the pair distribution
function.
\footnote{
Assuming that the wave-function and derivatives are regular when two coordinates do
not coincide and that they are zero at infinity, an estimate of the rate of decay for the off-diagonal
terms can be obtained by performing integration by parts (on the intervals of continuous differentiability), which shows that they are of
the order  $\mathcal{O}(1/k^6)$. The same type of reasoning can be applied in the case of
periodic  boundary conditions (no trapping potential) in a box of length $L$ by considering
periodic wave-functions and derivatives  and discrete momentum $k=2 \pi  s/L$ with $s$ an integer, see \cite{OD}.
}
This result shows
that the momentum distribution presents a $1/k^4$ decay with extensive contact
$C=(4/a^2)\int g^{(2)} (x,x)\, dx$. We see that the contact is expressed in
terms of the local pair distribution function a feature which is also present
in the multi-component case. Eq.~(\ref{i2}) (which was first derived in
\cite{OD}) is valid not only for an arbitrary pure state but also for any
statistical mixture (see Sect.~\ref{General}) if we replace $g^{(2)}(x,x)$ by
its thermal expectation value $g^{(2)}_T(x,x)$ .

\subsection{Two-component systems}\label{SS2}

Two-component systems already present all the features that will allow us to easily identify the general
structure of the contact for the $\kappa$-component case. There are three possibilities: two types of bosons,
two types of fermions and the Bose-Fermi mixture. If we assume that in the case of the Bose-Fermi mixture
 particles of type $1$ are bosons and particles of type $2$ are fermions the Hamiltonian density is
\be\label{hamYG}
\mathcal{H}(x)= \sum_{\sigma=\{1,2\}}\frac{\hbar^2}{2m_\sigma}\6_x\Psi^\dagger_\sigma(x)\6_x\Psi_\sigma(x)
+V_{i}(x)
+\sum_{\sigma=\{1,2\}}\left(V_\sigma(x)-\mu_\sigma\right)\Psi^\dagger_\sigma(x)\Psi_\sigma(x)\, ,
\ee
with $V_i(x)$ the interacting potential
\be
V_{i}(x)=\left\{\begin{array}{lr}
  \frac{g_{11}}{2} \Psi^\dagger_1(x)\Psi^\dagger_{1}(x)\Psi_{1}(x)\Psi_1(x)+
  \frac{g_{22}}{2} \Psi^\dagger_2(x)\Psi^\dagger_{2}(x)\Psi_{2}(x)\Psi_2(x)+
  g_{12} \Psi^\dagger_1(x)\Psi^\dagger_{2}(x)\Psi_{2}(x)\Psi_1(x) & \mbox{ BB case } \\
  g_{12} \Psi^\dagger_1(x)\Psi^\dagger_{2}(x)\Psi_{2}(x)\Psi_1(x)  & \mbox{ FF case } \\
  \frac{g_{11}}{2} \Psi^\dagger_1(x)\Psi^\dagger_{1}(x)\Psi_{1}(x)\Psi_1(x)+
  g_{12} \Psi^\dagger_1(x)\Psi^\dagger_{2}(x)\Psi_{2}(x)\Psi_1(x) & \mbox{ BF case }\\
\end{array}
\right.
\ee

The case of balanced fermions ($\mu_1=\mu_2$) at zero temperature was investigated by Barth and Zwerger
\cite{BZ} using operator product expansion techniques and constitutes the only 1D multi-component system
for which a comprehensive set of Tan relations was derived. The $1/k^4$ tail of the momentum distribution
for two-component impenetrable bosons and fermions was analytically derived from the Fredholm determinant
representation of the Green's function in \cite{CSZ}.

We will also assume that in the wave-function the coordinates of the particles of type $1$ are $x_1,\cdots,
x_{N_1}$ and the coordinates of particles of type $2$ are $x_{{N_1}+1},\cdots, x_{N_1+N_2}$. This also
implies that $I_1=\{1,\cdots,N_1\}$ and $I_2=\{N_1+1,\cdots,N_1+N_2\}$ with the total number of particles
$N=N_1+N_2$.  Remembering that for fermionic particles $g_{\sigma\sigma}^{(2)}(x,x)=0$ we can treat
all cases simultaneously. We will compute first the large momentum distribution of type 1 particles i.e,
$n_1(k)$. As in the Lieb-Liniger case we will focus on the large-k limit of the Fourier transform of the
wave-function. We introduce the following notation for the Fourier transform ($i\in I_1$)
\be
\tilde\psi_N^i(x_1,\cdots,\underset{\underset{i}{\uparrow}}{k},\cdots,x_N)=\int e^{-i k x}
\psi_N (x_1,\cdots,\underset{\underset{i}{\uparrow}}{x},\cdots,x_N)\, dx\, .
\ee
The main difference in analysing the Fourier transform of the two-component system compared with
the single component case is given by the fact that the discontinuity of the derivative is different
depending  on the  type of particles that collide.  Using Lemma \ref{lemma} and (\ref{disc}) we obtain
($\lim_{x_i\rightarrow x_j}X_{ij}=x_j$)
\be
\lim_{k\rightarrow\infty}\tilde\psi_N^i=
\sum_{\substack{j=1\\j\ne i}}^{N_1}\frac{2 e^{-i k x_j}}{a_{11}\, k^2}
\psi_N(x_1,\cdots,\underset{\underset{i}{\uparrow}}{x_j},\cdots,\underset{\underset{j}{\uparrow}}{x_j},
\cdots, x_N)+ \sum_{\substack{j=N_1+1\\ j\ne i}}^{N}\frac{2 e^{-i k x_j}}{a_{12}\, k^2}\psi_N(x_1,\cdots,
\underset{\underset{i}{\uparrow}}{x_j},\cdots,\underset{\underset{j}{\uparrow}}{x_j},\cdots, x_N)
\ee
where the first sum represents the contribution of the coinciding points $x=x_j$ with $x_j\in I_1$ and
the second sum gives the  contributions of $x=x_j$ with $x_j\in I_2$). Using this last result
we find
\begin{align}
\lim_{k\rightarrow\infty} \int&\prod_{\substack{j=1,\, j\ne i}}^N dx_j\left|\tilde\Psi_N^i(x_1,\cdots,
k,\cdots,x_N)\, dx\right|^2\nonumber\\
&=\int\prod_{\substack{j=1\\ j\ne i}}^N dx_j \left(
\sum_{\substack{j=1\\j\ne i}}^{N_1}\frac{4}{a_{11}^2 k^4} |\psi_N(\cdots,
\underset{\underset{i}{\uparrow}}{x_j},\cdots,\underset{\underset{j}{\uparrow}}{x_j},\cdots)|^2+
\sum_{\substack{j=N_1+1\\ j\ne i}}^{N}\frac{4}{a_{12}^2 k^4} |\psi_N(\cdots,
\underset{\underset{i}{\uparrow}}{x_j},\cdots,\underset{\underset{j}{\uparrow}}{x_j},\cdots|^2\, \right).
\end{align}
where again the off-diagonal terms vanish faster than $1/k^4$ due to the
Riemann-Lebesgue lemma. The momentum distribution (\ref{mom}) requires the
summation of all terms of this type with $i\in I_1$ with the result
\begin{align}
\lim_{k\rightarrow \infty} n_1(k)&\sim\frac{1}{k^4}\left(\frac{4}{a_{11}^2}\int g^{(2)}_{11}(x,x)\, dx+\frac{4}{a_{12}^2}
\int g^{(2)}_{12}(x,x)\, dx\right)\, .
\end{align}
In a similar fashion we obtain for $n_2(k)$
\begin{align}
\lim_{k\rightarrow \infty}n_2(k)&\sim\frac{1}{k^4}\left(\frac{4}{a_{22}^2 }\int g^{(2)}_{22}(x,x)\, dx+\frac{4}{a_{12}^2 }
\int g^{(2)}_{12}(x,x)\, dx\right)\, .
\end{align}
From these results we infer that $C_1=(4/a_{11}^2)\int g^{(2)}_{11}(x,x)\, dx+(4/a_{12}^2)\int g^{(2)}_{12}(x,x)
\, dx$ and $C_2=(4/a_{22}^2)\int g^{(2)}_{22}(x,x)\, dx+(4/a_{12}^2)\int g^{(2)}_{12}(x,x)\, dx$. In the case
of the two-component fermionic gas $g^{(2)}_{11}(x,x)=g^{(2)}_{22}(x,x)=0$ and therefore the contacts are equal
$C_1=C_2$ even in the case of an unbalanced system ($\mu_1\ne \mu_2$ or, equivalently, we can say that the
magnetic field is nonzero). In the case of the two-component bosonic gas the contacts for an arbitrary pure state
which is not the groundstate are equal only in the balanced system ($\mu_1= \mu_2$) when $g^{(2)}_{11}(x,x)=
g^{(2)}_{22}(x,x)$. The groundstate of two-component bosons is fully polarized \cite{EL,Suto,YL} which means that
at zero temperature $C_1=(4/a_{11}^2) \int g^{(2)}_{11}(x,x)\, dx$ and $C_2=0$.

\subsection{General case}\label{General}

Guided by the results of Sect.~\ref{SS2} we can now present the general
results for a $\kappa$-component system.  For all systems described by the
Hamiltonians (\ref{ham}) the momentum distribution of each type of particles
has the asymptotic behavior $\lim_{k\rightarrow \infty} n_\sigma(k)\sim
C_\sigma/k^4$ with the contact given by
\be\label{contactg}
C_\sigma=\sum_{\sigma'=1}^\kappa \frac{4}{a_{\sigma\sigma'}^2}\int g_{\sigma\sigma'}^{(2)}(x,x)\, dx=
\sum_{\sigma'=1}^\kappa \frac{4}{a_{\sigma\sigma'}^2}\int \langle \Psi_\sigma^\dagger(x)\Psi_{\sigma'}^\dagger(x)
\Psi_{\sigma'}(x)\Psi_\sigma(x)\rangle \, dx\, .
\ee
An alternative form of the previous expression involving derivatives of the energy with respect to the
scattering lengths can be derived using the Helmann-Feynman theorem. We obtain $\6 E/\6 a_{\sigma\sigma'}=
\hbar^2 \int g_{\sigma\sigma'}^{(2)}(x,x)\, dx/ m_{\sigma\sigma'}  a_{\sigma\sigma'}^2 (1+\delta_{\sigma\sigma'})$
which shows that (\ref{contactg}) can be also written as
\be\label{contactge}
C_\sigma=\frac{4}{\hbar^2}\sum_{\sigma=1}^\kappa m_{\sigma\sigma'}(1+\delta_{\sigma\sigma'})
\frac{\6 E}{\6 a_{\sigma\sigma'}}\, ,
\ee
where $\6 E/\6 a_{\sigma\sigma}=0$ if particles of type $\sigma$ are fermions. In the particular case of
two-component  fermions this last relation takes the form $\6 E/\6 a_{12}=\hbar^2 C_{1,2}/4 m_{12}$  which is also
known as the  adiabatic Tan theorem \cite{Tan1,BZ}.

Eqs.~(\ref{contactg}) and (\ref{contactge}) are valid for any pure state. In the case of an arbitrary statistical
mixture defined by $\langle \cdot \rangle=\sum p_n \langle \psi_n|\cdot |\psi_n\rangle$ the contact is  $C_\sigma
=\sum p_n C^{(n)}_\sigma$ with $C^{(n)}_\sigma= \sum_{\sigma'=1}^\kappa (4/a_{\sigma\sigma'}^2)\int
g^{(2) n}_{\sigma\sigma'}(x,x) \, dx\, $ and  $g^{(2) n}_{\sigma\sigma'}(x,x)=\langle \psi_n| \Psi_\sigma^\dagger(x)
\Psi_{\sigma'}^\dagger(x)\Psi_{\sigma'}(x)\Psi_\sigma(x)|\psi_n\rangle.$ In the particular case of the grand-canonical
ensemble Eq.~(\ref{contactge}) takes the form
\be\label{contactgegc}
C_\sigma=\frac{4}{\hbar^2}\sum_{\sigma=1}^\kappa m_{\sigma\sigma'}(1+\delta_{\sigma\sigma'})
\left(\frac{\6 \Phi}{\6 a_{\sigma\sigma'}}\right)_{\mu_\sigma,T}\, ,
\ee
with $\Phi(\mu_\sigma,T,a_{\sigma\sigma'})$ the grand-canonical potential.

Let us consider the particular case of systems which are characterized by
$m_\sigma=m_\sigma'=m$ and $a_{\sigma\sigma'} =a$ for all $\sigma, \sigma'
\in\{1,\cdots,\kappa\}$ and repulsive interactions $a<0$ (if $V_\sigma(x)=0$
these systems would be integrable but all the consideration below hold also in
he presence of an external potential). The total contact is now given by
$C_{tot}=(4/a^2)\sum_\sigma \sum_{\sigma'} \int g_{\sigma\sigma'}^{(2)}(x,x)\,
dx\, .$ The pair distribution function $g_{\sigma\sigma'}^{(2)}(x,x)$
quantifies the probability that two particles of type $\sigma$ and $\sigma'$
to be found in the same region of space.  If we consider a system comprised of
only fermionic particles then the pair distribution functions and the total
contact will be largest when the system is balanced i.e., when the number of
particles of each type is equal $N_\sigma=N_{\sigma'} =N/\kappa$. Switching a
magnetic field and keeping the total number of particles constant the
resulting imbalance will cause the pair distribution functions and total
contact to decrease. Therefore, for fixed density in a fermionic system the
total contact is largest in the balanced system and is a monotonously
decreasing function of the magnetic field vanishing when only one type of
particles survive. This behavior was numerically confirmed in the case of
two-component fermions at zero and finite temperature in \cite{PK} and for
finite numbers of particles in two- and three-component systems in
\cite{DJARMV1}.  If in the case of fermionic systems the ground-state is
antiferromagnetic as a consequence of the Lieb-Mattis theorem \cite{LM} in
purely bosonic systems the ground-state is ferromagnetic
\cite{EL,Suto,YL}. The wave-function of the ground-state is real and positive
$\Psi_{[J=N S]}^0(x_1,\cdots,x_N)\ge 0$ for all $x_1,\cdots,x_N$ (the
wave-function can be zero at coinciding points only in the case of infinite
repulsive interactions) with total spin $J=N S$ where $S$ is each boson's
spin.  An eigenstate with total spin $J<N S$ denoted by $\Psi_{[J<N S]}^0$ is
no longer positive \cite{YL} and numerical data for two-component bosons
reveal that \cite{PK1,PKF} $ \langle \Psi_{[J=N/2]}^0|V_{int}
|\Psi_{[J=N/2]}^0\rangle> \langle \Psi_{[J<N/2]}|V_{int} |\Psi_{[J<N/2]}
\rangle\, $ with $V_{int}$ the interaction term of the Hamiltonian
(\ref{ham}). The last relation also implies that (subscripts identify the
state) $ g_{11}^{(2)}(x,x)_{[J=N/2]}> \sum_{\sigma,\sigma'=1}^\kappa
g_{\sigma\sigma'}^{(2)}(x,x)_{[J<N/2]} $ which shows that the total contact of
the two-component bosonic system is largest in the ferromagnetic state. The
minimum (which is nonzero) is reached for the balanced system in contrast with
the fermionic case. It is sensible to assume that this behavior of the total
contact is also present in the case of $\kappa$-component systems with
$\kappa>2$.

\section{Short distance expansion of correlators}\label{S4}

The behavior of the wave-functions in the vicinity of the coinciding points can be used to derive  short
distance expansions of the pair distribution functions and one-body density matrices. In both cases
the first terms of these expansions can be expressed in terms of quantities that can be obtained from the
thermodynamic properties of the system.

\subsection{The pair distribution function}

We start with the short distance expansion  of  the pair distribution function  $g_{\sigma\sigma'}^{(2)}
(X+m_{\sigma\sigma'}x/m_\sigma, X-m_{\sigma\sigma'}x/m_{\sigma'})$ defined in (\ref{pair}). Using
Eq.~(\ref{disc}) and the definition we find
\begin{align}
g_{\sigma\sigma'}^{(2)}(X+\frac{m_{\sigma\sigma'}x}{m_\sigma},X-\frac{m_{\sigma\sigma'}x}{m_{\sigma'}})&
=\sum_{i\in I_\sigma}\sum_{j\in I_{\sigma'}}\int \prod_{k\ne\{i,j\}}\, dx_k
|\psi_N(x_1,\cdots,\underset{\underset{i}{\uparrow}}{X},\cdots,\underset{\underset{j}{\uparrow}}{X},
 \cdots,x_N)|^2 \left[1-\frac{2|x|}{a_{\sigma\sigma'}}+\mathcal{O}(|x|^2)\right]\, ,\nonumber\\
&=g_{\sigma\sigma'}^{(2)}(X,X)\left[1-\frac{2|x|}{a_{\sigma\sigma'}}+\mathcal{O}(|x|^2)\right]\, .
\end{align}
A simple application of the Helmann-Feynman theorem (see the remark before Eq.~(\ref{contactge})) allows
us to express the expansion of the spatially integrated distribution function as
\be
\int g_{\sigma\sigma'}^{(2)}(X+\frac{m_{\sigma\sigma'}x}{m_\sigma},X-\frac{m_{\sigma\sigma'}x}{m_{\sigma'}})
\, dX =a_{\sigma\sigma'}^2(1+\delta_{\sigma\sigma'})\frac{m_{\sigma\sigma'}}{\hbar^2}\frac{\6 E}
{\6 a_{\sigma\sigma'}} \left[1-\frac{2|x|}{a_{\sigma\sigma'}}+\mathcal{O}(|x|^2)\right]\, .
\ee
Previous expressions simplify considerably in the case of homogeneous systems
with particles of equal mass. In the case of two-component fermions it was
first shown in \cite{BZ} that the $|x|$ non-analyticity at
short distances imply that the static structure factor $S(k)\sim 1+ n \int dx\,
e^{-ik x} [g_{12}^{(2)}(X+x/2,X-x/2)-1]$ will decay at large momenta like
$S(k\rightarrow\infty)\sim 1-const\, C_{1}/k^2$ with $C_1$ the contact. In 3D
the tail of the static structure factor of spin-$\frac{1}{2}$ fermions is
proportional to $C_1/k$  from which the contact was measured using Bragg spectroscopy
\cite{KHLDV1,KHHDH}.

\subsection{The one-body density matrix}

Here, we derive the short distance asymptotics of the spatially integrated
one-body density matrix i.e, $G_{\sigma\sigma}^{(1)}(x)=\int
g_{\sigma\sigma}^{(1)}(X+x/2,X-x/2)\, dX$, which can also be understood as the
Fourier transform of the momentum distribution $G_{\sigma\sigma}^{(1)}(x)=\int
e^{i k x}n_\sigma(k)\, dk/2\pi.$ We will assume a short distance expansion of
the type \be
g_{\sigma\sigma}^{(1)}(X+x/2,X-x/2)=N_\sigma(X)+a_1^\sigma(X)x+a_2^\sigma(X)x^2+a_3^\sigma(X)|x|^3+\mathcal{O}(x^4)\,
.  \ee In the previous expansion we have not considered an $|x|$ term because
we already know that $n_\sigma(k)\sim C_\sigma/k^4$ which together with Lemma
\ref{lemma} means that the first nonanalytic term is $|x|^3$ from which we can
compute $\int a_3^\sigma(X)\, dX=C_\sigma/12$. We still need to determine
$a_1^\sigma(X)$ and $a_2^\sigma(X)$.  Expanding $e^{i kx}$ to second order we
obtain
\[
G_{\sigma\sigma}^{(1)}(x)\sim \int \left(1+ikx +\frac{(ik x)^2}{2!}\right)\frac{n_\sigma(k)}{2\pi}\ dk+\mathcal{O}(x^3)\, ,
\]
which allows to make the identifications $\int N_\sigma(X)\, dX=N_\sigma\, ,$
$\int a_1^\sigma(X)\, dX=i \int k\,n_\sigma (k)/2\pi\, dk$ and $\int
a_2^\sigma(X)\, dX=- \int k^2 n_\sigma (k)/4\pi\, dk.$ If we consider a system
with symmetric momentum distribution $n(k)=n(-k)$ then $\int
a_1^\sigma(X)=0$. Also it is easy to see that $\int a_2^\sigma(X)\,
dX=-m_\sigma T_\sigma/\hbar^2$ with $T_\sigma$ the kinetic energy of the
$\sigma$ type particles. Therefore, we obtain
\be\label{greene}
G_{\sigma\sigma}^{(1)}(x)\equiv\int g_{\sigma\sigma}^{(1)}(X+x/2,X-x/2)\, dX=N_\sigma-\frac{m_\sigma T_\sigma}
{\hbar^2} x^2+\frac{C_\sigma}{12}|x|^3
+\mathcal{O}(x^4)\, .
\ee
Similar short distance expansions of the Green's function were obtained in \cite{OD} for the Lieb-Liniger model and
in \cite{BZ}  for two-component fermions.

\section{Thermodynamic identities}\label{S5}

In this Section we are going to derive several universal thermodynamic identities connecting  the total energy,
trapping energy, momentum distribution and derivatives of thermodynamic potentials with respect to the scattering
lengths.

The total energy of the system $E=\langle H\rangle$ can be expressed as the sum of the kinetic energy $T$, interaction
energy $I$ and trapping energy $V$ i.e., $E=T+I+V$. Rewriting the kinetic energy of the Hamiltonian (\ref{ham}) in
momentum space and employing the Helmann-Feynman theorem we obtain
\be\label{e1}
E=\sum_\sigma \int  \frac{dk}{2\pi} \frac{\hbar^2 k^2}{2m_\sigma} n_\sigma(k)-\sum_{\sigma\le\sigma'}\frac{\6 E}
{\6 a_{\sigma\sigma'}} a_{\sigma\sigma'}+V
\ee
where  $\6 E/\6 a_{\sigma\sigma}=0$ if the particles
of type $\sigma$ are fermions. Eq.~(\ref{e1}) also holds in the case of an arbitrary statistical mixture  (see
the discussion in  Sect.~\ref{General}). In the case of the grand-canonical ensemble we only need to replace
$\6 E/\6 a_{\sigma\sigma}$ by $\6 \Phi/\6 a_{\sigma\sigma}$.

We will also derive two more universal relations known as the pressure and energy identities. As a preliminary step
we need to investigate certain scaling relations. The Hamiltonian in the first quantization for the general case is
given by (\ref{ham2}) with coupling strengths $g_{\sigma\sigma'}=-\hbar^2/(m_{\sigma\sigma'} a_{\sigma\sigma'})$ and
external potentials of the type $V_\sigma(x)=m_\sigma \alpha_\nu x^\nu$ with $\nu=2,4,6,\cdots$. In the case of the
harmonic trapping potential $\nu=2$ and $\alpha_2=\omega^2$ with $\omega$ the trapping frequency. We want to find
what is the behaviour of the Hamiltonian (\ref{ham2}) under a scaling transformation $x_i= x_i'/\lambda$ with $\lambda
\in\mathbb{R}^*_+.$  Under this transformation we have $d^2/dx^2\rightarrow\lambda^2 d^2/dx'^2,$ $\delta(x)\rightarrow
\delta(x'/\lambda)=\lambda \delta(x')$ and $V_\sigma(x)\rightarrow m_\sigma\alpha_\nu x'^\nu/\lambda^\nu$.
Therefore,
\be\label{i6}
H_{L/\lambda}\left(\{\lambda^{-1}a_{\sigma\sigma'}\},\lambda^{\nu+2}\alpha_\nu,\{\lambda^2 \mu_\sigma\}\right)=
\lambda^2 H_{L}\left(\{a_{\sigma\sigma'}\},\alpha_\nu,\{\mu_\sigma\}\right)\, ,
\ee
where we have introduced the notations $\{\lambda^{-1}a_{\sigma\sigma'}\}=\{\lambda^{-1}a_{11}, \lambda^{-1}a_{12}\cdots,
\lambda^{-1}a_{\kappa\kappa}\}\, ,\,$ $\{\lambda^{2}\mu_\sigma\}=\{\lambda^2\mu_1, \cdots,\,\lambda^2\mu_\kappa\}$ and
the subscripts $L/\lambda$ and $L$ denote the size of the finite systems. An obvious consequence of Eq.~(\ref{i6}) is
that  a similar relation exists between the energy eigenvalues of the two Hamiltonians. Using the definition of the
grand-canonical potential we obtain
\be\label{i7}
\Phi_{L/\lambda}\left(\{\lambda^{-1}a_{\sigma\sigma'}\},\lambda^{\nu+2}\alpha_\nu,\{\lambda^2 \mu_\sigma\},\lambda^2 T
\right)= \lambda^2 \Phi_{L}\left(\{a_{\sigma\sigma'}\},\alpha_\nu,\{\mu_\sigma\},T\right)\, .
\ee
In the case of an homogeneous system $(\alpha_\nu=0)$ introducing the grandcanonical potential per length $\phi=\Phi/L$
we have
\be\label{i8}
\phi\left(\{\lambda^{-1}a_{\sigma\sigma'}\},\{\lambda^2 \mu_\sigma\},\lambda^2 T\right)=
\lambda^3 \phi\left(\{a_{\sigma\sigma'}\},\{\mu_\sigma\},T\right)\, .
\ee

\subsection{Pressure identity for the homogeneous system}

Taking the thermodynamic limit and differentiating (\ref{i8}) with respect to $\lambda$ and then setting $\lambda=1$ we
obtain
$
-\sum_{\sigma\le \sigma'}\frac{\6 \phi}{\6 a_{\sigma\sigma'}}\, a_{\sigma\sigma'}+2 \sum_{\sigma} \frac{\6 \phi}{\6
\mu_\sigma}\, \mu_\sigma+ 2\frac{\6 \phi}{\6 T}\, T=3\phi\, .
$
From thermodynamics we have $\phi=-p=\mathcal{E}-T\mathcal{S}-\sum_\sigma \mu_\sigma n_\sigma$ where $p$ is the pressure,
$\mathcal{E}$ is the energy density, $\mathcal{S}$ the entropy density and $n_\sigma$ the particles densities. Also we
have $\frac{\6 \phi}{\6 T}=-\mathcal{S}$ and $\frac{\6 \phi}{\6 \mu_\sigma}=-n_\sigma$. Collecting everything we obtain the
pressure relation
\be\label{pressurei}
 p=2\mathcal{E}+\sum_{\sigma\le \sigma'}\frac{\6 \phi}{\6 a_{\sigma\sigma'}}\, a_{\sigma\sigma'}=2\mathcal{E}-\mathcal{I}\, ,
\ee
with $\mathcal{I}$ the interaction energy density.

\subsection{Energy identity for the inhomogeneous system}
Similarly as in the homogeneous case by taking the thermodynamic limit and differentiating (\ref{i7}) with respect to
$\lambda$  and then setting $\lambda=1$ we  obtain
$
-\sum_{\sigma\le \sigma'}\frac{\6 \Phi}{\6 a_{\sigma\sigma'}}\, a_{\sigma\sigma'}+ (2+\nu)\frac{\6 \Phi}{\6 \alpha_\nu}
\alpha_\nu+ 2 \sum_{\sigma} \frac{\6 \Phi}{\6 \mu_\sigma}\, \mu_\sigma+ 2\frac{\6 \Phi}{\6 T}\, T=2\Phi\, .
$
Using $\Phi=E-TS -\sum_\sigma \mu_\sigma N_\sigma$ and
$\frac{\6 \Phi}{\6 \alpha_\nu}\alpha_\nu=\langle \sum_\sigma m_\sigma\alpha_\nu x^\nu\rangle\equiv V$
with $V$ the trapping energy, we obtain the energy identity
\be\label{energyi}
 E=\frac{2+\nu}{2} V-\frac{1}{2}\sum_{\sigma\le \sigma'}\frac{\6 \Phi}{\6 a_{\sigma\sigma'}}\, a_{\sigma\sigma'}\,
 =\frac{2+\nu}{2} V+\frac{1}{2} I\, .
\ee

The universal identities (\ref{e1}), (\ref{pressurei}), and (\ref{energyi}) were first derived in the case of two-component
fermions in \cite{BZ} and in the case of the Lieb-Liniger model in \cite{V}.

\section{The integrable case}\label{S6}

The Hamiltonian (\ref{ham}) is integrable when $V_\sigma(x)=0$ and all masses and scattering lengths are
independently equal \cite{LL,G1,Y1,Suth}. In order to make contact with the relevant literature in this
section we are going to use units of $\hbar=2m=k_B=1$ and introduce $c=-2/a$. In this case the systems
are homogeneous
and it is useful to introduce the contact density denoted by  $\mathcal{C}_\sigma=C_{\sigma}/L$,  energy density
$\mathcal{E}=E/L$, etc. Unfortunately, compared with the nonintegrable case, from the thermodynamics of
the systems we can only obtain the total contact
\be
\mathcal{C}_{tot}\equiv\sum_{\sigma=1}^\kappa\mathcal{C}_\sigma=c^2\frac{\6 \mathcal{\phi}}{\6 c}\, ,
\ee
with $\phi$  the grandcanonical potential per length. If we would have considered  the ground state or any
other pure state $\mathcal{C}_{tot}=c^2\frac{\6 \mathcal{E}}{\6 c}$ with $\mathcal{E}$ the energy density of
the state. In the case of a balanced pure fermionic or bosonic system we can derive the individual contacts using
$\mathcal{C}_\sigma=c^2\frac{\6 \phi}{\6 c}/\kappa$. The short distance expansions of the correlators become
\be
g_{\sigma\sigma'}^{(2)}(X+x/2,X-x/2)=g_{\sigma\sigma'}^{(2)}(X,X)\left(1+c|x|+\mathcal{O}(|x|^2)\right)
\ee
and
\be
\sum_{\sigma=1}^\kappa g_{\sigma\sigma}^{(1)}(X+x/2,X-x/2)=n-\frac{ \mathcal{T}_{tot}}{2 } x^2+
\frac{\mathcal{C}_{tot}}{12}|x|^3 +\mathcal{O}(x^4)\, .
\ee
with $n$ the total density and  $\mathcal{T}_{tot}$ the total kinetic energy density. Again, in a balanced pure fermionic
or bosonic system we have $g_{\sigma\sigma}^{(1)}(X+x/2,X-x/2)=(n-\frac{ \mathcal{T}_{tot}}{2 } x^2+
\frac{\mathcal{C}_{tot}}{12}|x|^3)/\kappa$. The thermodynamic identities take the form
\be
p=2\mathcal{E}-\frac{\mathcal{C}_{tot}}{c}\, ,\ \ \ \ \ E=\frac{2+\nu}{2}V+\frac{1}{2}
\frac{\mathcal{C}_{tot}}{c}\, ,
\ee
where for the last relation we have considered the integrable system in the presence of the external potential
$V(x)=m\alpha_\nu x^\nu$ with $V=\langle m\alpha_\nu x^\nu\rangle$ the trapping energy.

\section{Conclusions}\label{S7}

In this paper we derived Tan relations for multi-component 1D models with
contact interactions. We considered the most general case in which the masses
and the inter- and intra-couplings can be different and also took into account
the presence of an external potential and magnetic field. We showed that the
tail of the momentum distribution presents a universal $1/k^4$ decay with the
amplitude given by the contact which can be expressed in terms of the local
pair distribution functions. In addition we obtained short distance expansions
of the correlation functions and several thermodynamic identities. We have
also argued that in the case of purely fermionic systems at fixed density the
total contact is a decreasing function of the imbalance while in a purely
bosonic system the converse is true. The case of systems with mixed symmetry
was numerically investigated by Decamp \textit{et al.} \cite{DJARMV2} (see
also \cite{DJARMV1,DAFAMV,PLHZC}) and it was found that the ground-state has
the most symmetric wave-function allowed by the mixture which can be
understood as a generalization of the Lieb-Mattis theorem \cite{LM}.

The dependence of the contact on the temperature is an issue that needs
further exploration. In the case of integrable systems at high-temperatures it
is expected that it is monotonously increasing as a function of the
temperature for all values of the coupling strengths
\cite{PK,PKF,DJARMV1,DJARMV2}. However, at low-temperatures and strong
coupling in the case of two-component models the total component exhibits a
local minimum which signals a significant momentum reconstruction. This
phenomenon is present for all integrable models with two-components
\cite{PK,PKF} and is due to the transition from the low-temperature phase to
the spin-incoherent regime \cite{CSZ}.  We expect that this behavior is also
present in the case of systems with more than two-components. This is left for
further investigation.

\acknowledgments

O.I.P acknowledges financial support from the LAPLAS 4 program of the Romanian
National Authority for Scientific Research (CNCS-UEFISCDI). A.K. acknowledges
financial support by the DFG in the framework of the research group FOR 2316.
O.I.P would like to thank the organizers of the workshop ``Correlations in Integrable
Quantum Many-Body Systems'', September 5-9, 2017, Hannover where part of this research
was performed.

\appendix

\section{Discontinuity of the derivative of the wave-function}\label{a1}

Here we prove Eq.~(\ref{disc}). Because we consider all the other coordinates kept fixed it will
be sufficient to investigate the nonanalyticity of the wave-function for a system with the Hamiltonian
\be\label{ah}
H=-\frac{\hbar^2}{2m_\sigma}\frac{\6^2}{\6 x_i^2}-\frac{\hbar^2}{2m_{\sigma'}}\frac{\6^2}{\6 x_j^2}+g_{\sigma\sigma'}\delta(x_i-x_j)\, ,
\ee
where we have neglected the external potential and chemical potential terms which are irrelevant for the present
discussion. Introducing the center of mass and relative coordinates of the two particles i.e.,
$X_{ij}= (m_\sigma x_i+m_{\sigma'}x_j)/(m_\sigma+m_{\sigma'})$ and  $x_{ij}=x_i-x_j$, the eigenvalue
problem of the Hamiltonian (\ref{ah}) becomes
\[
\left(-\frac{\hbar^2}{2(m_\sigma+m_{\sigma'})}\frac{\6^2}{\6 X_{ij}^2}-
\frac{\hbar^2}{2m_{\sigma\sigma'}}\frac{\6^2}{\6 x_{ij}^2}+g_{\sigma\sigma'}\delta(x_{ij})\right)\psi(X_{ij},x_{ij})
=E \psi(X_{ij},x_{ij})\, ,
\]
with $m_{\sigma\sigma'}=m_\sigma m_{\sigma'}/(m_\sigma+m_{\sigma'})$ the reduced mass of the two particles.
Integrating over $x_{ij}$ in a small vicinity of $0$ we obtain
\[
-\frac{\hbar^2}{2m_{\sigma\sigma'}}\int_{-\epsilon}^{+\epsilon}\frac{\6^2\psi(R_{ij},x_{ij})}{\6 x_{ij}^2}\,dx_{ij}
+g_{\sigma\sigma'}\int_{-\epsilon}^{+\epsilon}\delta(x_{ij})\psi(R_{ij},x_{ij})\, dx_{ij}=0\, .
\]
implying $\6_{x_{ij}}\psi(R_{ij},0_+)-\6_{x_{ij}}\psi(R_{ij},0_-)=2g_{\sigma\sigma'}m_{\sigma\sigma'}
\psi(R_{ij},0)/\hbar^2$ which  shows that the derivative of the wave-function is discontinuous. We can write
$\Psi(X_{ij},x_{ij})=\psi(X_{ij},0)\left(1-|x_{ij}|/a_{\sigma\sigma'}+\mathcal{O}(x_{ij}^2)\right)$ where we have used
$g_{\sigma\sigma'}=-\hbar^2/ (m_{\sigma\sigma'} a_{\sigma\sigma'})$. Using $x_i=X_{ij}+m_{\sigma\sigma'}
 x_{ij}/m_\sigma$ and $x_j=X_{ij}-m_{\sigma\sigma'} x_{ij}/m_\sigma'$ the generalization to the case of $N$ particles
is given by Eq.~(\ref{disc}).

\end{document}